\documentclass[11pt]{article}
\usepackage{hyperref} 
\newcommand{\tab}{\hspace*{2em}}
\begin{document} 
\pdfoutput=1 

\title{Impinging Jets and Droplet Dynamics} 
\author{Xiaodong Chen, Prashant Khare, Dongjun Ma, and Vigor Yang \\ 
\\\vspace{6pt} School of Aerospace 
Engineering, \\ Georgia Institute of Technology, Atlanta, GA 30332, USA} 
\maketitle 
\section*{Abstract}
        \tab In this fluid dynamics video, results from high fidelity numerical simulations are presented, which have been carried out to study the flow and droplet dynamics of liquid sheets formed by two impinging jets. A three-dimensional Volume-of-Fluid (VOF) method with adaptive mesh refinement (AMR) based on octree meshes [1] is used to simulate the various flow patterns associated with impinging jets, secondary breakup and binary collision of droplets. In addition to AMR, a thickness based refinement algorithm is also developed and implemented to efficiently resolve the various scales of surface tension driven interfacial flows.

	Oblique collision of two cylindrical, laminar jets causes the liquid to flow outward from the impact point, creating a thin sheet which lies in a plane perpendicular to the plane containing the two jets. This sheet eventually disintegrates into ligaments and/or droplets. Breakup of the liquid sheet is dominated by viscosity and surface tension effects (Reynolds and Weber number). In this video, impinging water jets are investigated, and different flow patterns, such as closed rim, open rim, rimless and ligament, are identified and studied. Depending on the operating conditions, droplets are shed circumferentially from the periphery of the sheet in some cases, whereas in others, ligaments are fragmented from the leading edge of the sheet, which further breaks down into droplets following the Rayleigh mechanism. The periodic waves from the point of impingement are clearly seen on the surface of the sheet. The impact waves cause the early breakdown of the sheet downstream of the impingement point, whereas waves amplified by aerodynamic stresses control the breakdown of the rest of the sheet and the ligaments. 

	Detailed understanding of droplet dynamics, which includes equal and unequal sized droplet collision, bouncing, coalescence, separation, mixing, secondary breakup and mass transfer, is also obtained systematically at a wide range of operating conditions. Underlying processes governing droplet dynamics are investigated by analyzing the overall energy budget. In addition, an advanced visualization technique using the Ray-tracing methodology is implemented to gain direct insight into the detailed physics of droplet interaction. Theories are also established to predict the droplet behavior after collision, which shows good agreement with existing literature. A detailed account of the above mentioned phenomena can be found in [2] and [3].  

\section*{Acknowledgements}
This work was sponsored by the US Army Research Office under the Multi-University Research Initiative (MURI) under contract number W911NF-08-1-0124. The support and encouragement provided by Dr. Ralph Anthenien is greatly acknowledged. The authors are also thankful to Dr. Stephane Popinet for allowing them to use his VOF and AMR algorithms.

\section*{References}
[1] S. Popinet. An accurate adaptive solver for surface-tension-driven interfacial flows. Journal of Computational Physics, Vol. 228, No. 16, 2009, pp. 5838-5866.
\newline
[2] D. Ma, X. Chen, P. Khare, and V. Yang. Atomization and breakup characteristics of liquid sheets formed by two impinging jets. 49th AIAA Aerospace Sciences Meeting, January 2011, Orlando, Florida.
\newline
[3] X. Chen, D. Ma, P. Khare, and V. Yang. Detailed physics of binary droplet collision. 7th US National Technical Meeting of the Combustion Institute, March 2011, Atlanta, Georgia.  
\clearpage
\section*{Submission Contents}
Two videos are contained in this submission for the American Physical Society - 64th DFD Meeting, Gallery of Fluid Motion. 
\begin{enumerate} 
\item ImpingingJetsDropletDynamics\_HR.mpg - High resolution video approprite for DFD 2011 Gallery of Fluid Motion. 
\item ImpingingJetsDropletDynamics\_LR.mpg - Low resolution video appropriate for web viewing. 
\end{enumerate} 
\end{document}